\documentclass[aps,nofootinbib,twocolumn]{revtex4}
\usepackage{graphicx}
\usepackage{amsfonts}
\usepackage{amsmath, amsthm, amssymb}
\usepackage{wasysym}
\begin{document}

\title{The Lyth Bound and the End of Inflation}

\author{Richard Easther} \email{richard.easther@yale.edu}
\affiliation{Department of Physics, Yale
University, New Haven  CT 06520, USA}
\author{William H.\ Kinney} \email{whkinney@buffalo.edu}
\author{Brian A. Powell} \email{bapowell@buffalo.edu}
\affiliation{Dept. of Physics, University at Buffalo,
        the State University of New York, Buffalo, NY 14260-1500}
\date{\today}

\begin{abstract}
\noindent 
We derive an extended version of the well-known Lyth Bound on the total
variation of the inflaton field, incorporating higher order
corrections in slow roll. We connect the field variation $\Delta\phi$ to
both the spectral index of scalar perturbations and the amplitude of
tensor modes.  We then investigate the implications of this bound for
``small field'' potentials, where the field rolls off a local
maximum of the potential. The total field variation during
inflation is {\em generically} of order $m_{\rm Pl}$, even for potentials
with a suppressed tensor/scalar ratio. Much of the total field excursion arises in the last e-fold of inflation and in single field models this problem can
only be avoided  via fine-tuning or the imposition of a symmetry.  Finally,  we discuss
the implications of this result for inflationary model building in string theory and
supergravity. 
\end{abstract}

\pacs{98.80.Cq}

\maketitle

\section{Introduction}

Inflation \cite{Guth:1980zm} has proven to be a fantastically successful paradigm for explaining why the universe is so big, so old, and so flat, and provides a mechanism for generating the primordial cosmological perturbations. 
 However, a fundamental description of the physics responsible for inflation remains elusive: no convincing model for inflation yet exists. We can, however, discern the broad characteristics of the  inflationary epoch from observational data. The absence of a detectable gravitational wave contribution to the CMB (Cosmic Microwave Background) anisotropies tells us that the energy density of the universe during inflation was something less than $10^{16}\ {\rm GeV}$. The fact that the primordial perturbation spectrum is close to scale invariant tells us that the potential function of the field or fields responsible for inflation (the {\it inflaton}) must be very flat. Finally, the highly suppressed amplitude of the perturbation spectrum ($\delta \sim 10^{-5}$) requires the introduction of a small parameter to the inflationary Lagrangian, either as a small self-coupling or as a hierarchy of mass scales. 
 
Probably the best guess as  to the fundamental origin of the inflationary
potential is that the inflaton (if there is only one) is an
effective degree of freedom in the low-energy  limit of quantum gravity \cite{Lyth:1998xn}.
The last few years have seen substantial progress toward realizing this
hypothesis within string theory \cite{Dvali:1998pa,Kachru:2003sx,Silverstein:2003hf,Burgess:2004kv,Alishahiha:2004eh,Burgess:2005sb,Dimopoulos:2005ac,Aazami:2005jf,Easther:2005zr}, but no dominant
and compelling stringy model of inflation has emerged, and string
phenomenology is far from fully understood.  In this context, effective
field theory is a  powerful tool.   Since the inflaton potential is
tightly constrained by data, if the inflaton field takes on a vacuum expectation value (vev) at
which higher order operators contribute significantly to the potential,
it is very unlikely that the delicate balance between the height and
slope of the potential will survive intact.  
Therefore effective field theory arguments favor a vev for the inflaton which is small in Planck units: $\Delta \phi \ll m_{\rm Pl}$. Lyth \cite{Lyth:1996im} derived a lower bound on the variation in the inflaton field during inflation in terms of the ratio $r$ between tensor and scalar perturbations generated during inflation, known as the {\em Lyth Bound}. Combined with a theoretical prejudice based on effective field theory, the Lyth Bound predicts a  strong suppression of primordial gravitational waves relative to the observed amplitude of the scalar perturbations.

The significance of the Lyth Bound has been the subject of much debate. In particular, Linde has argued that only the {\em energy density} during inflation must be sub-Planckian and that field values greater than $m_{\rm Pl}$ are physically consistent \cite{Linde:2004kg}, but this argument does not apply to generic supergravity or superstring inspired models \cite{Easther:2005zr}. 
Lyth's argument is restricted to single field models, and multifield models can evade it in two ways. Firstly, one might imagine that several separate inflationary epochs are concatenated together.\footnote{See \cite{Easther:2004ir,Burgess:2005sb} for specific proposals along these lines, and \cite{Aazami:2005jf} for a more general argument about the implementation of multifield models in the string landscape.}   Secondly, the large number of scalar fields introduced by the string landscape makes it natural to look for implementations of {\em assisted inflation\/} \cite{Liddle:1998jc} where a large number of similar (if not identical) scalar fields act cooperatively to drive inflation.  The resulting cosmology mimics that produced by a single field with a large vev, but the individual fields never take on trans-Planckian expectation values.  A concrete example of this approach is provided by  ``N-flation'' \cite{Dimopoulos:2005ac,Easther:2005zr}. 

Questions surrounding the robustness and generality of the Lyth Bound will only grow in importance as measurements of the perturbation spectrum put ever tighter constraints on the amplitude of primordial  tensor fluctuations \cite{Kinney:2003gi,Alabidi:2005qi}. In this paper, we do not attempt to resolve the debate about the self-consistency of effective field theory in inflation, but simply pose the question: If we suppose that future observations show a strongly suppressed amplitude of primordial tensor fluctuations, what will that tell us about the inflaton potential? 

To investigate this question, we derive an extended version of the Lyth Bound accurate to higher order in slow roll. This allows us to connect the field variation $\Delta\phi$ to  the spectral index of scalar perturbations, as well as to the amplitude of tensor modes.
We then investigate the implications of this bound for ``small field'' potentials, where the field rolls off a local maximum of the potential. We find that, in the absence of fine-tuning, the field variation during the inflationary period is {\em generically} of order $m_{\rm Pl}$, even for models with a very small tensor/scalar ratio.  Specifically,  near the end of inflation, the field variation is rapid enough to ensure that  $\Delta\phi \sim m_{\rm Pl}$ during the last e-fold of expansion. This is true even when the field variation is small during the epoch in which the observable perturbations are generated. We show that this ``last e-fold problem'' can only be solved by suppressing the inflaton mass term -- either by fine-tuning or the imposition of a symmetry. The remainder of the paper is structured as follows: Section \ref{sec:extendedlythbound} derives the higher-order extension of the Lyth Bound. Section \ref{sec:smallfieldmodels} discusses the issue of field variation in generic small-field models and the ``last e-fold'' problem. Section \ref{sec:conclusions} contains a summary and conclusions.

\section{The Extended Lyth Bound}
\label{sec:extendedlythbound}
Inflationary cosmology predicts the existence of a spectrum of
large-scale tensor fluctuations (gravitational waves) in addition to
scalar fluctuations (adiabatic density perturbations).  Both are expected
to
contribute to the measured temperature anisotropy and polarization of the
cosmic microwave
background (CMB), and different inflationary models predict varying
contributions.  The relative contributions are
quantified in terms of the tensor-to-scalar ratio
\begin{equation}
\label{r}
r = \frac{P_T}{P_{\mathcal R}} = 16\epsilon,
\end{equation}
where
\begin{eqnarray}
P_{T}^{1/2} &&= \frac{4 H}{ m_{\rm Pl} \sqrt{\pi}} \\
P_{\mathcal R}^{1/2} &&= \frac{H}{m_{\rm Pl} \sqrt{\pi \epsilon}} \label{R}
\end{eqnarray}
are the amplitudes of tensor and scalar perturbations, respectively, and
\(\epsilon\) is the first Hubble slow-roll parameter
\begin{equation}
\label{eps}
\epsilon \equiv \frac{m_{\rm Pl}^2}{4\pi}\left(\frac{H'(\phi)}{H(\phi)}\right)^2.
\end{equation}
There is much interest in the possible detection of a tensor
contribution to the CMB temperature anisotropy.
Upcoming CMB missions will have vastly increased sensitivity to the B-mode polarization signal characteristic of gravitational radiation. The Planck satellite \cite{planck} will have the sensitivity to detect $r \sim 0.05$ and successors such as NASA's Inflation Probe \cite{inflationprobe} may achieve a sensitivity of $r \sim 0.01$ or better \cite{Kinney:1998md,Verde:2005ff}. Direct detection of the primordial gravitational wave background with the envisioned Big Bang Observer mission \cite{BBO} has a projected sensitivity of $r \sim 10^{-3} - 10^{-4}$ \cite{Ungarelli:2005qb,Kinney:2005in}. 
Together with the COBE normalization, \(\delta_H \simeq 1.9 \times 10^{-5}\),
a detection of tensor fluctuations will (assuming a nearly scale-invariant spectrum) fix
the energy scale of inflation via the relation
\begin{equation}
\Lambda \sim r^{1/4} \times (3.3 \times 10^{16}\, {\rm GeV}),
\end{equation}
where \(\Lambda^4 = V\) is the height of the inflationary potential
when scales of cosmological interest exited the horizon. Furthermore,
since \(r \sim \epsilon\) and \(\epsilon \propto (V'/V)^2\) to lowest order in
slow roll, a measurement of tensor modes will also reveal information
about \(V'\) when slow roll is valid.  Detection of the
B-mode will therefore constrain the shape of the inflaton potential
enough to eliminate many currently viable
inflationary models.  The discovery of primordial gravity waves is
therefore a
valuable and exciting prospect. However, there are theoretical arguments
which suggest that the amplitude of the tensor component is expected to  be well
below current observational limits.

Lyth \cite{Lyth:1996im} showed that the amplitude of the tensor component
(and therefore \(r\)) depends on the total inflaton field variation,
\(\Delta \phi\), during the time that observable scales exit the horizon.
From the expression
\begin{equation}
\label{phiN}
\frac{d\phi}{dN} = \frac{m_{\rm Pl}}{2\sqrt{\pi}}\sqrt{\epsilon},
\end{equation}
and by using Eq. (\ref{r}), it is possible to
write the field variation as a function of \(r\),
\begin{equation}
\Delta \phi = \frac{m_{\rm Pl}}{8\sqrt{\pi}}\sqrt{r}|\Delta N|,
\end{equation}
where \(N\) is the number of e-folds before the end
of inflation.  It is assumed that there is negligible variation in
\(r\) over the period \(|\Delta N|\), which is a good approximation to
lowest order in slow roll.   In his original paper, Lyth considered
\(\Delta \phi\)
as scales
corresponding to \(1 < l \apprle 100\) were exiting the horizon, during
which
time the universe expands by an amount \(|\Delta N| \simeq 4\). Since this gives a lower limit on the total field variation, the result is an inequality known as the {\em Lyth Bound}:\footnote{The addition of other theoretical priors can result in a strengthened bound \cite{Boubekeur:2005zm}.}
\begin{equation}
\label{lythbound}
\Delta \phi \apprge m_{\rm Pl}\sqrt{\frac{r}{4\pi}}.
\end{equation}
This indicates that for sizeable values of \(r\), the inflaton
field must roll a distance of order \(m_{\rm Pl}\) while observable scales
are exiting the horizon.  Field variations
\(\apprge m_{\rm Pl}\) are, in fact,  ubiquitous in inflationary
model-building, and
essential for the success of chaotic inflation and certain models
based on symmetry
breaking potentials.
Such a large field variation may, however, be problematic from the
standpoint of effective field
theory if the underlying inflationary model is embedded in either
supergravity or string theory.

It is widely believed that inflation can be appropriately described by
the evolution of a
fundamental scalar field: a field existing in the low energy limit of
some more fundamental theory
such as supergravity or string theory.  It is possible to describe these
fundamental theories at
low energies by an effective field theory, obtained by integrating out
high momentum degrees of freedom and rescaling fields and coupling
constants. An effective field theory description of inflation thus
involves an effective potential for the inflaton of the form
\begin{equation}
\label{veff}
V(\phi) = V_0  + \frac{1}{2}m^2\phi^2 + \phi^4 \sum_{p=0}^{\infty}
\lambda_p
\left(\frac{\phi}{m_{\rm Pl}}\right)^p.
\end{equation}
The terms of order \(p > 0\) have couplings of negative mass dimension,
and are therefore nonrenormalizable.
However, provided \(\phi \ll m_{\rm Pl}\), the nonrenormalizable terms
remain under control and the series is at least asymptotically convergent.
Therefore, if we hope to be able to make observable predictions based on
an effective field theory, we need \(\Delta \phi < m_{\rm Pl}\) as the  scales relevant to these predictions are leaving the horizon.  This is a conservative estimate: while we work with the full Planck mass, the appropriate vertex factor for a loop expansion in gravity  is actually the reduced Planck mass, $m_{\rm Pl}/\sqrt{8 \pi}$, about a factor of five smaller.    Ensuring that this relationship holds can easily lower the amplitude of the tensor modes to the point where they are effectively undetectable.  This strong theoretical prior on the inflationary model space is a stringent constraint on a key
inflationary observable and we now ask:  can the 
Lyth Bound can be extended to constrain  other observables?  

The power spectrum of superhorizon-sized perturbations (\(k \ll aH\)) is conventionally expressed  in terms of the comoving wavenumber, \(k\),
\begin{equation}
P_{\mathcal R}^{1/2} \propto k^{n-1},
\end{equation}
where \(n\) is the spectral index.  When slow roll is valid, the spectral
index can be written\footnote{Note that the algebraic form of the equation below differs slightly from that often encountered in the potential slow roll formalism.}
\begin{equation}
\label{n}
n = 1 - 4\epsilon + 2\eta,
\end{equation}
where \(\eta\) is the second Hubble slow-roll parameter,
\begin{equation}
\label{eta}
\eta \equiv \frac{m_{\rm Pl}^2}{4\pi}\left(\frac{H''(\phi)}{H(\phi)}\right).
\end{equation}
We wish to obtain \(\Delta \phi\) as a function of \(\eta\),
so we will extend the Lyth Bound to next order in slow roll.  Consider the
Taylor expansion of the
inflaton field \(\phi(N)\),
\begin{equation}
\label{exp}
\phi(N) = \phi_0 + \frac{d\phi}{dN}(\Delta N) +
\frac{1}{2}\frac{d^2\phi}{dN^2}(\Delta N)^2 +
\cdots,
\end{equation}
where the coefficients in the series are given by the flow equations,
\begin{eqnarray}
\frac{d\phi}{dN} &=& \frac{m_{\rm Pl}}{2\sqrt{\pi}}\sqrt{\epsilon} \nonumber \\ &=& \frac{m_{\rm Pl}}{\sqrt{4 \pi}} \frac{H'}{H}, \\
\frac{d^2\phi}{dN^2} &=&  \frac{d}{dN} \frac{d\phi}{dN} \nonumber \\ &=& \frac{m_{\rm Pl}}{2\sqrt{\pi}}\sqrt{\epsilon}(\eta -
\epsilon).
\end{eqnarray}
Eq. (\ref{exp}) then becomes
\begin{equation}
\label{newbound}
\Delta \phi = \frac{m_{\rm Pl}}{2\sqrt{\pi}}\sqrt{\epsilon}|\Delta N|\left[1
+ \frac{1}{2}(\eta -
\epsilon)(\Delta N)\right].
\end{equation}
The first term in Eq. (\ref{newbound}) is Lyth's original result and the
additional term
provides an \(\eta\)-dependent correction.  Using Eqs. (\ref{r},\ref{n})
and taking \(\Delta N
\simeq -4\) we obtain the extended Lyth Bound \footnote{Note that \(N\)
decreases as one goes forward in time.
In this paper we
adopt the convention \(\sqrt{\epsilon} \propto H'(\phi)\) so that
\(\dot{\phi} \propto
-\sqrt{\epsilon}\).}
\begin{equation}
\label{extended}
\frac{\Delta \phi}{m_{\rm Pl}} \apprge \sqrt{\frac{r}{4\pi}}\left[1 - (n-1)
-\frac{r}{8}\right].
\end{equation}
The effect of the extra term in Eq. (\ref{newbound}) depends on the sign
of \(\eta\).  We are
particularly interested in models for which  \(\eta < 0\), as they often lead
to a
detectable value of \(r\)  \cite{Dodelson:1997hr}.  These consist of
potentials that arise from
spontaneous symmetry breaking, pseudo Nambu-Goldstone bosons (natural
inflation) \cite{Freese:1990rb,Adams:1992bn,Arkani-Hamed:2003wu,Arkani-Hamed:2003mz,Freese:2004un,Blanco-Pillado:2004ns,Brax:2005jv},  and certain  chaotic inflation
scenarios.  For \(\eta
< 0\), the bound   tightens
as \(|\eta|\) becomes
large.  During slow roll, \(\eta\) can be given in terms of the
inflaton potential \(V(\phi)\):
\begin{equation}
\eta = \frac{m_{\rm Pl}^2}{8\pi}\left[\frac{V''(\phi)}{V(\phi)} -
\frac{1}{2}\left(\frac{V'(\phi)}{V(\phi)}\right)^2\right]. 
\end{equation}
For a given value of \(\epsilon\) and vacuum energy density
\(\Lambda\), \(\eta\) grows more negative
as the curvature of the
potential increases over the region defined by \(\Delta N\).  The field \(\phi\)
rolls more quickly as the  concavity of the potential increases, and thus travels further 
as the universe expands by \(|\Delta N|\) e-folds.
For a given \(r\), the extended Lyth Bound Eq. (\ref{extended}) is
tightened slightly for
models with red spectra (\(n<1\))
and loosened for those with blue spectra (\(n>1\)).

\section{The last e-fold}
\label{sec:smallfieldmodels}

Equation (\ref{extended}) determines the field variation \(\Delta \phi\) as scales with \(1 < l \apprle 100\)
leave the horizon. Lyth \cite{Lyth:1996im} pointed out that the lower order version of this result, Eq. (\ref{lythbound}), is necessarily a lower limit on the total field variation. It is conservative, as  $\epsilon$  can increase significantly in the remaining \(N\) e-folds of inflation,
resulting in \(\Delta \phi \gg m_{\rm Pl}\).\footnote{Efstathiou and Mack \cite{Efstathiou:2005tq} use a
stochastic approach to extend the
Lyth Bound over the entire inflationary period.  For models satisfying
the spectral index constraint \(0.92 < n < 1.06\), they obtain the bound
\begin{equation} \frac{\Delta \phi}{m_{\rm Pl}} \approx 6r^{1/4}
\end{equation} where \(\Delta \phi\) is the variation over the remaining 55 e-folds of
inflation.} For example, if  \(V \sim \phi^4\),
 \(\Delta \phi \simeq 4m_{\rm Pl}\) over the last \(50-60\)
e-folds of inflation.  These models predict \(r \simeq
0.1-0.2\), so \(\Delta \phi \sim 0.1m_{\rm Pl}\)
over the first  \(4\) e-folds of inflation. We thus see that while this interval corresponds to around 7\% of the total expansion of the universe, it provides only \(2.5\%\) of the field variation.

If inflation is driven by a single scalar field whose potential has the form of  Eq.
(\ref{veff}), we need to ensure that 
\(\Delta \phi\) does not exceed \(m_{\rm Pl}\) over the full course of inflation, and not just over the interval during which currently observable scales are leaving the horizon. For a given \(N\) we can write down the equivalent field value, \(\phi_N\). For a specific potential, \(\phi_N\) is a function  of
\(\phi_e\), the field value at the end of inflation which is determined from the slope of the potential.  As we now
show, field variations during the {\em last} e-fold of
inflation are generically of order \(m_{\rm Pl}\), making this constraint
difficult to satisfy.

We begin with an example.  Consider the potential
\begin{equation}
\label{chaotic}
V(\phi) = \Lambda^4\left(\frac{\phi}{\mu}\right)^p,
\end{equation}
where \(\Lambda^4 \propto V_0\) is the ``height'' of the potential
and \(\mu\) is the ``width''.  Such potentials are characteristic of
chaotic inflation scenarios \cite{Linde:1983gd} and typically result in
\(\Delta \phi > m_{\rm Pl}\) during inflation.  
To lowest order in slow roll,
\begin{equation}
\label{epssr}
\epsilon(\phi) =
\frac{m_{\rm Pl}^2}{16\pi}\left(\frac{V'(\phi)}{V(\phi)}\right)^2,
\end{equation}
which for the case of Eq. (\ref{chaotic}) becomes
\begin{equation}
\label{chaoticeps}
\epsilon(\phi) = \frac{m^2_{Pl}}{16\pi}\left(\frac{p}{\phi}\right)^2.
\end{equation}
From Eq. (\ref{phiN}), we obtain \(\phi\) as a function $N$,
\begin{equation}
\label{int}
\phi^2_e - \phi_N^2 = -\frac{m_{\rm Pl}^2}{4\pi}pN .
\end{equation}
We find \(\phi_e\) by
setting \(\epsilon(\phi_e) = 1\) in Eq. (\ref{chaoticeps})\footnote{The
reader may object to the use of \(\epsilon\) written to lowest order in
slow roll Eq. (\ref{epssr}) to obtain a field value when slow roll has
clearly broken down.  Use of the exact expression for \(\epsilon\) Eq.
(\ref{eps}) actually yields a {\em larger} field variation than that
obtained in Eq. (\ref{27}), but agreement between the two curves is within 5\%.}
\begin{equation}
\label{phiend}
\phi_e = \frac{m_{\rm Pl}}{4\sqrt{\pi}}p.
\end{equation}
We have now fully specified \(\phi_N\), and if \(p=4\) the field
variation over the {\em last} e-fold is
\begin{equation}
\label{27}
\Delta \phi = |\phi_e - \phi_1| = \frac{m_{\rm Pl}}{2\sqrt{\pi}}(\sqrt{8}-2)
\approx
0.8\frac{m_{\rm Pl}}{2\sqrt{\pi}}.
\end{equation}
This is much larger than  the field variation over the period when cosmologically relevant perturbations are generated, but yields only a single e-fold of inflation.  Moreover, the field variation grows with \(p\). Now consider  
\begin{equation}
\epsilon(N) = \frac{p}{p+4N},
\end{equation}
giving \(\epsilon(1) = 1/2\) for \(p=4\). We can understand the general case by noting that in Eq. (\ref{phiN}), if \(\epsilon\)
does not vary considerably as the universe expands by \(|\Delta N|\) e-folds, we can write
\begin{equation}
\label{approx}
\Delta \phi \approx \frac{m_{\rm Pl}}{2\sqrt{\pi}}\sqrt{\epsilon} |\Delta N|.
\end{equation}
Since inflation ends when \(\epsilon =1\), during the final e-fold 
\(\sqrt{\epsilon}|\Delta N| \sim 1\), resulting in a field
variation
\(\Delta \phi \sim m_{\rm Pl}/2\sqrt{\pi}\).  
As \(\Delta \phi\) is the integral of \(\sqrt{\epsilon}\) over
the last e-fold, the area under the curve in the figure 
is proportional to \(\Delta \phi\).  When \(V(\phi)
\propto \phi^4\), \(\epsilon\) is close to unity when $N=1$, and is exactly unity (by  definition) when \(N = 0\).  

Eq. (\ref{phiN}) is a general relationship between \(\phi\) and
\(N\), so the conclusion that a slowly varying \(\epsilon(N)\) leads to
\(\Delta \phi\) of order \(m_{\rm Pl}\) over the last e-fold is actually model
independent.  Consequently,  there is a large class of
inflationary potentials with this property, even though the explicit calculation here assumes  \(V(\phi) \propto \phi^p\).  

In order to keep the field variation small and solve this ``last e-fold'' problem, we need a potential for which $\epsilon$ rises very rapidly to unity as $N$ approaches zero. Due to
the vast number of inflation models consistent with current constraints
\cite{Peiris:2003ff,Barger:2003ym,Hannestad:2001nu,Kinney:2003uw,Seljak:2004xh}, rather than investigating $\Delta \phi$ for the last e-fold of inflation on a case-by-case
basis, we turn to {\em Monte Carlo reconstruction\/}  \cite{Easther:2002rw}, which relies on stochastic evaluation of the inflationary flow equations \cite{Hoffman:2000ue,Kinney:2002qn}. By selecting initial conditions
in the inflationary parameter space and integrating the flow equations,
we can generate a vast collection of models, and then filter out those which satisfy any specific 
constraints.  Monte Carlo reconstruction relies on a truncated slow-roll hierarchy, so it effectively assumes that the potential contains no sharp kinks or corners, since  even very high order derivatives of $H(\phi)$ and $V(\phi)$ are non-zero in the vicinity of such a feature. However, we are excluding this class of potentials by hypothesis when we assume that the end of inflation is due to a smooth transition to $\epsilon = 1$ from $\epsilon \sim 0$, rather than an almost instantaneous jump, so our analysis here is entirely self-consistent.   
\begin{figure}
\centerline{\includegraphics[angle=270,width=3.5in]{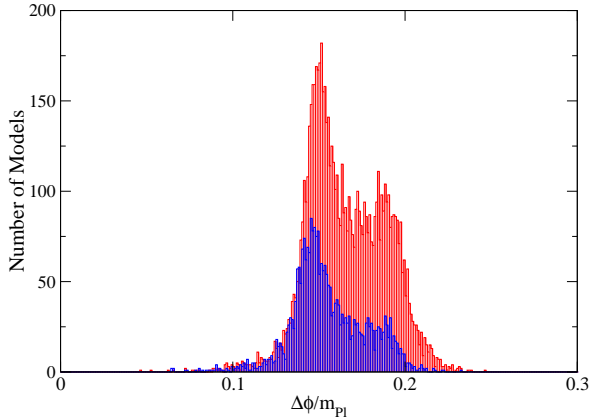}}
\caption{Histogram of \(\Delta \phi/m_{\rm Pl}\) over the last e-fold for 12,000 inflation models
satisfying the constraint  \(0.9 < n < 1.1\) on the spectral index. Red ($n < 1$) and blue ($n > 1$) spectra are plotted separately.}
\label{figure2}
\end{figure}

 Having solved the flow equations, one may then 
recover the inflaton potential for the models that survive the filtering process \cite{Easther:2002rw}.
The results of such an analysis are presented in Figure \ref{figure2}.
The histogram displays \(\Delta \phi/m_{\rm Pl}\)
during the last e-fold of inflation for 12,000 models in a sixth-order flow reconstruction with spectral index
within \(0.9
< n < 1.1\) at astrophysical scales. The models at the left hand end of the histogram are those for which we might expect that higher order corrections to the inflaton potential are ignorable throughout the last 60 e-folds of inflation, and we see that they are  are extremely rare.   This result strongly suggests that \(\Delta \phi / m_{\rm Pl} \sim 1\) over the last e-fold for a generic inflationary model.  

All of the models
included in Figure \ref{figure2} are ``nontrivial'' models, for which inflation ends dynamically by $\epsilon$ passing through unity. These have potentials of either the {\em large field} or {\em small field} type.  
Large field models are characterized by a
scalar field initially displaced from the minimum of the potential by an
amount typically of order the Planck mass.  Chaotic inflation
scenarios \cite{Linde:1983gd}
arising from polynomial potentials, as well those arising from
exponential potentials \cite{Lucchin:1984yf},
are examples of large field models.
Since large field models generally require an initial
field value \(\phi_0 > m_{\rm Pl}\), these are already problematic from an 
effective field theory perspective.  
In contrast, small field
models are distinguished by a scalar field initially very close to a
maximum of the potential, and roll off the ``hill-top'' as inflation progresses.
(We use the term ``small field'' in the original sense of Ref. \cite{Dodelson:1997hr}: the {\em initial} field value must be small, but the field variation $\Delta\phi$ need not be.)
Examples include Coleman-Weinberg
(``new'' inflation) and pseudo Nambu-Goldstone bosons (natural inflation)  potentials.   We are interested in identifying which small field models lead to inflation with \(\Delta \phi \ll
m_{\rm Pl}\).  As we will find, the behavior of \(\epsilon(N)\) during the last e-fold will be particularly important.  In Fig. \ref{figure2} we see models with both red ($n < 1$)and blue ($n > 1$) spectra. Naively, a blue spectrum is associated with hybrid inflation, where inflation ends via an instability in a direction orthogonal to the inflaton trajectory.  However, the ``blue'' models here are correlated with a significant running, and  the spectral index is thus a function of scale. This blue to red transition is facilitated by including higher order terms in the slow roll expansion.   This breakdown of the standard model-classification at higher order is discussed in detail in Ref. \cite{Kinney:2005in}. For the analytic models we consider below, such higher-order effects are negligible, and expressions to first-order in slow roll are sufficient to describe the physics. 

Scalar fields with spontaneous symmetry
breaking potentials have long been seen as candidate inflationary models.   Such potentials possess a point of unstable equilibrium with non-vanishing vacuum energy (the false vacuum), and a point of stable equilibrium with effectively zero vacuum energy (the true vacuum).  If \(\phi\) is initially in (or near) the false vacuum, thermal and/or quantum fluctuations eventually cause the field to roll towards the true vacuum. These models  are known as small
field models, since \(\phi_0 \approx 0\) initially.  

During inflation, the field is near the top of the potential, \(\phi \ll \mu\), so we can assume
\begin{equation}
V(\phi) = \Lambda^4\left[1 - \frac{1}{p}\left(\frac{\phi}{\mu}\right)^p +
\cdots\right].
\end{equation}
Since inflation ends and slow roll breaks down
before the field reaches the true vacuum \(\phi \sim \mu\),
a realistic potential with
\(\Delta \phi \ll m_{\rm Pl}\) during inflation must have \(\mu
\ll m_{\rm Pl}\), while satisfying observational constraints on the spectrum for scales exiting
the horizon around \(N = 60\).  From the Lyth Bound, it is clear that
the  tensor-to-scalar ratio  is negligible at these scales.
Therefore, a spectral index consistent with a nearly scale-invariant
spectrum will be the key inflationary observable.  Models of inflation
occurring at low scales have been investigated in Refs. \cite{Kinney:1995cc,Knox:1992iy}, among others.

We will focus on the potential
\begin{equation}
\label{hill}
V(\phi) = \Lambda^4\left[1 -
\frac{1}{p}\left(\frac{\phi}{\mu}\right)^p\right],
\end{equation}
for \(p=2\) and \(p \geq 4\), and we will see that these two choices lead to significantly different inflationary phenomenology. Working to first order in the
slow roll parameters is sufficient for our purpose here, and we drop higher-order 
corrections. 
We start with \(p=2\), the so-called ``inverted
quadratic'' potential. This is  the generic case, since in the absence of a symmetry that suppresses the mass, the quadratic term will {\em always} dominate for small enough $\phi$. Many of the following results are derived in more detail in Refs. \cite{Kinney:1995cc,Kinney:1998md}. A similar, more recent analysis can be found in Ref. \cite{Alabidi:2005qi}.
The first slow-roll parameter, \(\epsilon(\phi)\), is
\begin{equation}
\label{30}
\epsilon(\phi) \simeq
\frac{1}{16\pi}\left(\frac{m_{\rm Pl}}{\mu}\right)^2\left(\frac{\phi}{\mu}\right)^2.
\end{equation}
From Eq. (\ref{phiN}), the number of e-folds before the end of inflation
can be found as a function of \(\phi\),
\begin{equation}
\label{31}
N(\phi) \simeq 8\pi\left(\frac{\mu}{m_{\rm Pl}}\right)^2{\rm
ln}(\phi_e/\phi).
\end{equation}
Next, we determine the field value at \(N=60\), \(\phi_{60}\), as
this will be needed to calculate observables:
\begin{equation}
\frac{\phi_{60}}{\mu} = \sqrt{16\pi}\left(\frac{\mu}{m_{\rm Pl}}\right){\rm
exp}\left[-\frac{15}{2\pi}\left(\frac{m_{\rm Pl}}{\mu}\right)^2\right].
\end{equation}
Using Eq. (\ref{n}), we obtain the scalar spectral index at
\(N=60\),
\begin{align}
\label{2n}
n = 1 - 4\epsilon(\phi_{60}) + 2\eta(\phi_{60}) &\simeq 1 +
\frac{m^2_{Pl}}{4\pi}\left(\frac{V''(\phi_{60})}{V(\phi_{60})}\right)
\notag\\
&\simeq 1 - \frac{1}{4\pi}\left(\frac{m_{\rm Pl}}{\mu}\right)^2, 
\end{align}
where the approximation follows because the field is slowly rolling while
scales corresponding to \(\phi_{60}\) are leaving the horizon.  
We see that with \(\mu \ll m_{\rm Pl}\), the spectral index \(n\ll1\), resulting in an
extremely red spectrum.  Requiring \(n > 0.9\) yields  \(\mu \apprge 0.9m_{\rm Pl}\).  A scale
invariant spectrum, \(n = 1\), is achieved in the limit \(\mu
\rightarrow \infty\).  

Under what circumstances is it possible to satisfy the condition \(\Delta \phi \ll
m_{\rm Pl}\) with an inverted quadratic potential? In the limit $\phi \ll \mu$ the second slow-roll parameter becomes a constant,
\begin{equation}
\eta \simeq -\frac{1}{8 \pi} \left({m_{\rm Pl} \over \mu}\right)^2,\ \phi \ll \mu.
\end{equation}
Again, we find $\phi_e$, the field value at the end of inflation, by solving  \(\epsilon \equiv 1$ 
\begin{equation}
{\phi_e \over m_{\rm Pl}} = {1 \over 2 \sqrt{\pi} \left\vert\eta\right\vert}.
\end{equation}
Using Eq. (\ref{31}), we then obtain
\begin{equation} \label{phirun}
{\phi\left(N\right) \over m_{\rm Pl}} = {1 \over 2 \sqrt{\pi} \left\vert\eta\right\vert} e^{-\eta N}.
\end{equation}
Then $\phi_{60} \ll \phi_e$, and the total field variation during inflation is
\begin{equation}
{\Delta \phi \over m_{\rm Pl}} \simeq {\phi_e \over m_{\rm Pl}} = {1 \over 2 \sqrt{\pi} \left\vert\eta\right\vert}.
\end{equation}
The condition $\Delta\phi \ll m_{\rm Pl}$ is a {\em lower bound} on the magnitude of $\eta$,
\begin{equation}
\left\vert\eta\right\vert \gg {1 \over 2 \sqrt{\pi}}.
\end{equation}
Alternatively, we can regard this as a new manifestation of the $\eta$ problem that is endemic to inflationary models embedded in supergravity where (nearly) flat directions necessarily acquire a mass roughly equivalent to the Hubble parameter $H$, and $\eta \sim 1$ \cite{Copeland:1994vg}. This gives a constraint on  the spectral index,
\begin{equation}
n = 1 + 2 \eta \ll 0.4.
\end{equation}
Note that this is exactly the opposite of what one might expect from the extended Lyth Bound (\ref{extended}) where, for fixed $r$, we find that it is easier to achieve $\Delta\phi \ll m_{\rm Pl}$  in the limit $n \rightarrow 1$.
To understand this apparent discrepancy, we use Eqs. (\ref{30}) and (\ref{31}), to obtain \(\epsilon(N)\) 
\begin{equation}
\epsilon(N) = {\rm
exp}\left[-\frac{N}{4\pi}\left(\frac{m_{\rm Pl}}{\mu}\right)^2\right] = e^{-2 \left\vert\eta\right\vert N}.
\end{equation}
Thus we cannot naively take the small-$\eta$ limit with $r$ fixed. The rate of change of the field can be obtained from  Eq. (\ref{phiN}),
\begin{equation}
{d \phi \over d N} = {m_{\rm Pl} \over 2 \sqrt{\pi}} e^{-\left\vert \eta\right\vert N}.
\end{equation}
The change in the value of $\phi$ between $N_1$ and $N_2 < N_1$ is
\begin{equation}
{\Delta\phi(N_1 \rightarrow N_2) \over m_{\rm Pl}} = {1 \over 2 \sqrt{\pi} \left\vert\eta\right\vert} \left(e^{-\left\vert\eta\right\vert N_2} - e^{-\left\vert\eta\right\vert N_1}\right).
\end{equation}
Taking $n > 0.9$, so that $\eta \sim 0.05$, the field variation from $N = 60$ to $N = 56$ is then
\begin{equation}
\Delta\phi(60 \rightarrow 56) = 0.05 m_{\rm Pl},
\end{equation}
which satisfies $\Delta\phi \ll m_{\rm Pl}$. However, the field variation during the last e-fold is large:
\begin{equation}
\Delta\phi(1 \rightarrow 0) = 0.2 m_{\rm Pl}.
\end{equation}
As in our earlier flow analysis, virtually all of the field evolution happens near the {\em end} of inflation. This is not surprising, since the end of inflation is precisely where slow roll is breaking down. Therefore, while a small value of $r$ is a necessary condition for $\Delta\phi$ to be small in Planck units throughout inflation, it is not a {\em sufficient} condition. For any potential dominated by the inflaton mass term during inflation, the total field variation is always of order $m_{\rm Pl}$, even if the field variation is small early in inflation. 

Potentials for which the mass term is suppressed, either through a fine-tuning or through the imposition of a symmetry \cite{Kinney:1995cc}, can avoid the ``last e-fold'' problem. 
Consider Eq. (\ref{hill}) for \(p \geq 4\).  
Proceeding as in the \(p=2\) case, we find
\begin{equation}
\label{38}
\epsilon(\phi) \simeq
\frac{1}{16\pi}\left(\frac{m_{\rm Pl}}{\mu}\right)^2\left(\frac{\phi}{\mu}\right)^{2(p-1)}.
\end{equation}
The number of e-folds before the end of inflation is
\begin{equation}
\label{Np}
N(\phi) \simeq
8\pi\left(\frac{\mu}{m_{\rm Pl}}\right)^2\left(\frac{1}{p-2}\right)\left[\left(\frac{\mu}{\phi}\right)^{p-2}-\left(\frac{\mu}{\phi_e}\right)^{p-2}\right],
\end{equation}
and from \(N(\phi)\), \(\phi_{60}\) is
\begin{equation}
\label{39}
\frac{\phi_{60}}{\mu} \simeq
\left[\frac{2\pi}{15(p-2)}\left(\frac{\mu}{m_{\rm Pl}}\right)^2\right]^{1/(p-2)}.
\end{equation}
This approximation follows from Eq. (\ref{Np}) since the first term in
square brackets dominates when \(\phi \ll \mu\). 
From Eq. (\ref{n}), the spectral index evaluated at \(\phi_{60}\) is
\begin{equation}
\label{40}
n \simeq 1 - \left(\frac{1}{30}\right)\frac{p-1}{p-2},
\end{equation}
and  depends only on the order of the dominant term in the
potential Eq. (\ref{hill}).  This is in sharp contrast to the result for \(p=2\)
Eq. (\ref{2n}), 
where the spectral index is a function of 
\(\mu\).   The origin of the difference in
the form of \(n\) in Eqs. (\ref{2n}) and (\ref{40}) is that  \(V''\) is independent of \(\phi\) when
\(p=2\).  Since \(n -
1 \propto V''/V \sim -1/\mu^2\), as \(\mu\) decreases, \(n\) is driven
smaller regardless of where on the potential it is measured ({\em i.e.}
the value of \(\phi_{60}\)).  Potentials with \(p \geq 4\) have
\(V''/V \propto -1/\mu^2(\phi /\mu)^{p-2}\) which, for a given \(\phi\),
also becomes more negative as \(\mu\) decreases.  However, the point at
which we measure the spectral index, \(\phi_{60}\), also scales with
\(\mu\) as in Eq. (\ref{39}).  The scaling behavior
of \(V''/V\) and that of \(\phi_{60}/\mu\) exactly cancel, so that as we
go to smaller scales, the
spectral index is being measured at point on the potential for which
\(V''/V\) does not differ from one scale to the next.  

It is important to determine what additional constraints, if any, are
imposed on the scale \(\mu\) by the COBE normalization, 
\(\delta_H \simeq 1.9 \times 10^{-5}\), which is related to the amplitude of
scalar perturbations Eq. (\ref{R}) by \(\delta_H =
(2/5)P^{1/2}_{\mathcal R}\).  At lowest order, the perturbation
amplitude at \(\phi_{60}\) is given  in terms of the potential  by
\begin{equation}
\label{34}
P^{1/2}_{\mathcal R} \approx
\sqrt{\frac{32 \pi}{3}}\frac{V^{3/2}(\phi_{60})}{m^3_{Pl}V'(\phi_{60})}.
\end{equation}
The COBE normalization relates the vacuum energy density \(\Lambda^4\)
and the symmetry breaking scale \(\mu\) at \(\phi_{60}\),
\begin{equation}
\label{35}
\left(\frac{\Lambda}{\mu}\right)^2 = P^{1/2}_{\mathcal
R}\sqrt{\frac{3}{32 \pi}}\left(\frac{2\pi}{15(p-2)}\right)^{\frac{p-1}{p-2}}\left(\frac
{m_{\rm Pl}}{\mu}\right)^{\frac{p-4}{p-2}}.
\end{equation}
When \(p=4\), the
hierarchy  \(m_{\rm Pl}/\mu\) drops out and  \(\Lambda\)
scales linearly with \(\mu\).  This is expected, since the
\(\phi^4\) field theory is scale-free.  If one assumes perfectly
efficient reheating, it is possible to derive a lower bound on
\(\Lambda^4\).  In order to preserve standard hot big bang cosmology, inflation must occur above the nucleosynthesis scale, or \(\Lambda =  
100\,{\rm MeV} \sim 10^{-20}m_{\rm Pl}\).  From Eq. (\ref{35}), this places a lower
bound of \(\mu \apprge 1\, {\rm GeV}\) on the symmetry breaking scale.  
Therefore, the \(p=4\) potential is capable of driving
inflation at very low scales while providing an acceptable reheat
temperature. 
In general, \(\Lambda \sim
\mu(m_{\rm Pl}/\mu)^{(p-4)/(2p-4)}\) so that for a given
scale \(\mu<m_{\rm Pl}\), higher order potentials will admit larger
\(\Lambda\).  

\begin{figure}
\centerline{\includegraphics[angle=270,width=3.5in]{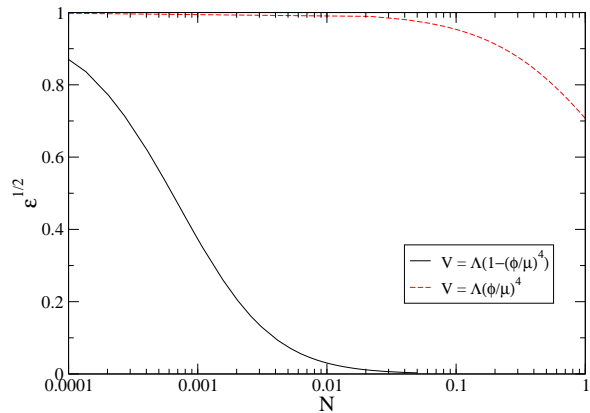}}
\caption{\(\sqrt{\epsilon}\) vs. \(N\) over the last e-fold for the potentials \(V(\phi) \ \lambda \phi^4\) (dotted line) and  \(V(\phi) =\Lambda^4[1-(\phi/\mu)^4]\) (solid line).  The
symmetry breaking scale for the inverted potential is \(\mu = 0.1m_{\rm Pl}\).}
\label{figure3}
\end{figure}

Lastly, it remains to determine the behavior of \(\epsilon(N)\) during
the last e-fold.  Using Eqs. (\ref{38}) and (\ref{Np}), we find 
\(\epsilon\) as a function of \(N\),
\begin{eqnarray}
\epsilon(N) &=&
\frac{1}{16\pi}\left[\frac{(p-2)N}{8\pi}\left(\frac{m_{\rm Pl}}{\mu}\right)^{\frac{3p-5}{p-2}}+ \right. \nonumber \\
 && \qquad \left.  \left(\frac{1}{16\pi}\right)^{\frac{p-2}{2p-2}}\right]^{-\frac{2p-2}{p-2}}.
\end{eqnarray}
For an illustrative comparison , consider the plot of \(\sqrt{\epsilon}\) vs.
\(N\) for potentials of the form \(V(\phi) = \lambda \phi^4$ and  \(V(\phi) = \Lambda^4[1-(\phi/\mu)^4]\), Fig. \ref{figure3}.
The symmetry breaking scale for the inverted potential is \(\mu = 0.1m_{\rm Pl}\), a fairly large value
considering that these potentials can lead to successful inflation at
much lower scales. The rapid falloff of $\sqrt{\epsilon}$ as a function of $N$ is exactly the behavior
expected out of a potential which satisfies \(\Delta \phi \ll m_{\rm Pl}\).

\section{Conclusions}
\label{sec:conclusions}

We consider the consequences of requiring  the  change in the inflaton field during its evolution  to be  sub-Planckian, or, $\Delta\phi < m_{\rm Pl}$. We derive a higher order version of Lyth's relationship between 
the field variation and the tensor/scalar ratio, $r$. This is valid to second order in slow roll, and is given by Eq.~(\ref{extended}). In addition to $r$, the field variation also depends on the scalar spectral index $n$.
Naively interpreted, this result appears to indicate that a spectrum close to scale invariance ($n \simeq 1$) makes it easier to ensure that $\Delta\phi$ is small, but we find that this is a model-dependent statement.

Arguments based on effective field theory suggest that 
a self-consistent inflation model must have $\Delta\phi < m_{\rm Pl}$ during the interval in which cosmological perturbations are laid down. To satisfy
this constraint with a single field model, the tensor/scalar ratio is
strongly suppressed: $r \ll 1$.  This argument amounts to putting a
strong theoretical prior on the inflationary model space -- namely that
we are working with a model derived from supergravity or string theory.
In this case, an effective field theory description of the inflaton
potential is very unlikely to be reliable when the inflaton acquires a
trans-Planckian vev, as the delicate balance between the height and slope of the potential required for successful inflation will almost certainly be disrupted. 

In this analysis, we do not take a position on whether it is reasonable to restrict the inflationary model space in this way. Rather, we take the view that observations will eventually either determine the value of $r$ or place a tight upper bound on its value, and we investigate the theoretical implications of these different possibilities. Lyth's  original discussion  considered only the field variation during the interval during which perturbations relevant to the lowest multipole moments CMB were being laid down,
relevant to the detectable signal from gravitational waves. 
We extend this result in two ways. Firstly, we include next-to-leading order terms in the slow-roll expansion, which yields an expression for $\Delta \phi$ which depends on both $r$, and the scalar spectral index $n$.  Secondly,  we consider the field variation over the final 60 e-foldings of inflation which gives a tighter constraint on the overall model space. In particular, we found that the field point generically evolves by an amount close to $m_{\rm Pl}$ during the final moments of inflation, leading to what we have dubbed the ``last e-fold problem'': almost all {\em minimally coupled, single field\/} inflationary models will have $\Delta  \phi \apprge m_{\rm Pl}$.

The simplest way to build an inflationary model with $\Delta\phi \ll m_{\rm Pl}$ is hybrid inflation \cite{Linde:1993cn}. In a hybrid scenario, the field evolves toward a minimum with nonzero vacuum energy.
On such a potential, the inflationary solution asymptotes to $\dot\phi =
0$ as $\phi$ approaches the minimum of the potential, after an {\em
infinite} amount of inflation \cite{Kinney:2005vj}.  Consequently, in these models inflation ends because of an instability in an orthogonal direction, induced by an auxiliary field. Clearly, this scenario yields an arbitrarily large amount of inflation with an arbitrarily small field variation. Such models are therefore well-behaved from the standpoint of effective field theory.  It might be argued that hybrid inflation is fine-tuned,  as it relies on having an auxiliary field with an instability exponentially close to the origin. In practice, hybrid models exploit the presence of a hierarchy of mass-scales in nature, and so do not involve any {\em extra\/} tunings.  

In the case of ``inverted'' potentials characteristic of spontaneous symmetry breaking, inflation ends dynamically when the slow roll parameter $\epsilon$ passes through unity, and no extra parameters are required to fix the endpoint of inflation. Near the origin, the potential is dominated by the lowest order term in its Taylor series, so we consider ``small-field'' models where $V(\phi) \sim 1 - (\phi /\mu)^p$.

Without tuning or a symmetry, the inflaton mass term dominates, and $p = 2$. In this case, we derive an expression for the field variation over the entire inflationary period, Eq~(\ref{phirun}), or $\Delta \phi / m_{\rm Pl}  \simeq {1 / (2 \sqrt{\pi} \left\vert\eta\right\vert)} \sim 1/(1-n)$.
Therefore as we approach the scale-invariant limit, $n \rightarrow 1$, the field variation becomes large in Planck units. Interestingly, most of the field evolution happens near the {\em end} of inflation, long after observable perturbations have been generated. The field variation can be small during the first four e-folds of inflation,
\begin{equation}
\Delta\phi(60 \rightarrow 56) = 0.05 m_{\rm Pl},
\end{equation}
thereby nominally satisfying $\Delta\phi \ll m_{\rm Pl}$.  The field variation is, however, {\em large} during the last e-fold,
\begin{equation}
\Delta\phi(1 \rightarrow 0) = 0.2 m_{\rm Pl}.
\end{equation}
Therefore, such models still suffer from the  ``last e-fold'' problem, and generically yield large field variations near the end of inflation. This behavior can be seen in the approximate flow relation
\begin{equation}
\Delta \phi \approx \frac{m_{\rm Pl}}{2\sqrt{\pi}}\sqrt{\epsilon} |\Delta N|,
\end{equation}
valid when $\epsilon \sim {\rm constant.}$ Near the end of inflation, $\epsilon \simeq 1$, and the field variation is generically of order $m_{\rm Pl}$, even for models which generate a negligible tensor amplitude on observable scales. This analytic result is consistent with conclusions based on a stochastic analysis using the flow equations, discussed here and in Ref. \cite{Efstathiou:2005tq}. Whether or not a field variation $\Delta\phi \sim 0.2 m_{\rm Pl}$ is actually a problem is a model-dependent statement \cite{Boyanovsky:2005pw}. One of us (WHK) has argued that field variations of this magnitude are perfectly acceptable in the context of natural inflation \cite{Freese:2004un}. Other models might incorporate such field variations in a natural way (see, for example, Ref. \cite{Berkooz:2005sf}). 

In order for the field variation during inflation to be very small in Planck units, the mass term for the inflaton must be suppressed. This might be accomplished by fine-tuning: for example a very flat potential with a sharp ``drop off'', so the potential resembles a mesa rather than  a rounded hilltop.  In this case, a Taylor expansion of the potential near the end of inflation is dominated by higher-order terms. (A similar tuning problem was identified in Ref. \cite{Boyle:2005ug} using a counting argument on a generic dimensionless paremeter set.) A less contrived alternative is to introduce a symmetry which forces the mass term to vanish, so that the potential is dominated by quartic (or higher-order) interactions during the inflationary period. In this case,  $\epsilon$ varies rapidly near the end of inflation, and $\Delta\phi \ll m_{\rm Pl}$ over the entire inflationary period, without the need for further tuning.  In Ref. \cite{Kinney:1995cc} WHK and Mahanthappa constructed a detailed model based on an explicitly broken local symmetry, demonstrating that one can achieve the desired suppression of the inflaton mass term within a realistic Lagrangian. We argue that this model satisfies the ``challenge'' issued by Boyle, Steinhardt, and Turok in Ref. \cite{Boyle:2005ug} to construct a deterministic, complete inflationary model which is consistent with $r < 10^{-2}$. The model is technically natural: no fine-tuning is required even when loop corrections are included, and the potential 
\begin{eqnarray}
V\left(\theta\right)&=& {3 v^4 \over 64 \pi^2} g^4 \times \cr
&& \left\lbrace \sin^4\left(\theta \over v\right)
\ln\left[g^2 \sin^2\left(\theta \over v\right)\right] - \ln\left(g^2\right)\right\rbrace \quad
\end{eqnarray}
has a smooth minimum, bounded below. The inflaton can be coupled to fermionic degrees of freedom, allowing reheating, without spoiling the form of the inflationary potential. A discussion of how to achieve a similar suppression of the mass term in extra-dimensional theories can be found in Ref. \cite{Arkani-Hamed:2003mz}. We therefore conclude that 
tuning arguments {\em cannot} be used as suggested by Boyle, {\it et al.} to place a lower-bound on the tensor/scalar ratio for inflation models. 

Conversely, there is no watertight argument that can be used to rule out inflationary models with $\Delta \phi > m_{\rm Pl}$, and the value of $r$ will ultimately be determined observationally. However, the analysis here shows that in typical models of inflation, the variation in the field value during the last e-fold of inflation can be significantly larger than the change during the epoch in which cosmological perturbations are laid down. Consequently, if one is considering a class of models for which $\phi \sim m_{\rm Pl}$ is disfavored, the analysis here can be used to put tight constraints on the models' parameter space.

Observationally, we are making rapid strides. Forthcoming observations will be capable of detecting a tiny deviation from scale-invariance. 
In the absence of running, a blue spectrum, ($n > 1$) will strongly favor hybrid-type inflation models. 
A red spectrum ($n<1$) favors ``inverted'' potentials typical of small-field inflation.\footnote{The WMAP3 data \cite{Spergel:2006hy}, released after the first version of this paper was written, strongly favor a red spectrum.} Current upper bounds on the tensor amplitude are much weaker than the ultimate theoretical limits on their detectability. If tensor modes are detected by future measurements, we will acquire a wealth of phenomenological information about the inflationary epoch \cite{Lidsey:1995np,Easther:2002rw}. If the tensor/scalar ratio turns out to be unobservably small, we can still learn much about inflation. However, despite recent progress, there is no guarantee that the theoretical questions surrounding the ``naturalness'' of inflation will be quickly resolved.

\section*{Acknowledgments}
RE and WHK thank the Aspen Center for Physics, where this project was initiated, for its hospitality and for providing a stimulating environment. We thank Eugene Lim, David Lyth and Liam McAllister for comments on a draft version of this paper, and Daniel Chung for helpful conversations.  RE is supported in part by the United States Department of Energy, under contract DOE-FC02-92ER-40704. WHK is supported in part by the National Science Foundation under grant NSF-PHY-0456777.

\end{document}